\documentclass[runningheads]{casc}

\usepackage{epsf}

\begin{document}
\pagestyle{headings}

\mainmatter
\title{
{\normalsize
\hfill SFB/CPP-04-27\\
\hfill TTP04-15\\
}
ParFORM: Parallel Version of the Symbolic Manipulation Program FORM }

\titlerunning{ParFORM}
\author{
M.~Tentyukov\inst{1}${}^,$\inst{2}
\and  D.~Fliegner\inst{3}
\and M.~Frank
\and A.~Onischenko\inst{4}
\and  A.~R\'etey
\and H.M.~Staudenmaier\inst{2}
\and J.A.M.~Vermaseren\inst{5}
}

\authorrunning{Tentyukov et al}

\institute{
BLTP, Joint Institute for Nuclear Research, Dubna, Russia \and
Institut   f\"ur Theoretische   Teilchenphysik,  Universit\"at
Karlsruhe, Germany \and Max-Planck-Institut f\"ur  Str\"omungsforschung, G\"ottingen,
Germany \and Department of Physics and Astronomy, Wayne State University,
Detroit, USA \and NIKHEF, Amsterdam}

\maketitle

\begin{abstract}
The symbolic manipulation program FORM is specialized to handle very
large algebraic expressions.
Some specific features of its internal structure
make FORM very well suited for parallelization.

After an introduction to the sequential version of FORM and the
mechanisms behind, we report on the status of our project
of parallelization. We have now a parallel version of FORM
running on Cluster- and SMP-architectures. This version can be used
to run arbitrary FORM programs in parallel.
\end{abstract}

\section{Introduction}

FORM \cite{form} is a program for symbolic manipulation of algebraic
expressions specialized to handle very large expressions of millions of
terms in an efficient and reliable way. That is why it is widely used in
Quantum Field Theory, where the calculation of the order of several
hundred (sometimes thousands) of Feynman diagrams is required. Currently
the actual version of FORM is called FORM3.

In context with this goal an improvement of efficiency is very important.
 Parallelization
is one of the most efficient ways to
increase performance.
So the idea to parallelize FORM is quite natural.

This paper reports the present status of our project
of FORM parallelization and the result is called ParFORM.

The main goal of parallel processing is to reduce wall-clock
time\footnote{The elapsed time from start to finish of a process.} i.e.
the user's waiting time. Parallelism does not come for free; it always has
some overhead with respect to serial execution, but it can significantly
reduce the wall-clock time.

Not every problem can be divided into parallel tasks. An example of a
parallelizable problem is the multiplication of two matrices. An example
of a non-parallelizable problem is the calculation of the Fibonacci series
(1,1,2,3,5,{\ldots}) by means of the recurrence formula
$F(k+2)=F(k+1)+F(k)$.

The last example has to be taken with  caution. It does not imply that every
recursion is necessarily non-parallelizable. For example, in the field
of perturbative calculation, where FORM has become a standard tool,
we often exploit recurrence relations. The latter, however, are
special
in a sense, that they are applied to every separate term in one expression
and
they are examples of local operations.

Of course, ParFORM cannot handle non-parallelizable problems,
but ParFORM is quite natural to apply it to any kind of parallelizable
problem. So, if we have to solve a parallelizable problem
which essentially involves local operations, the interior
structure of FORM
permits us to parallelize each step of a process.

There are some internal mechanisms of FORM that become
important in its parallel version and this will be described in the
next section.

\section{The Sequential Version of FORM}

\begin{figure}[ht]
\begin{center}
\vbox{\epsfxsize=0.82\linewidth \epsfbox{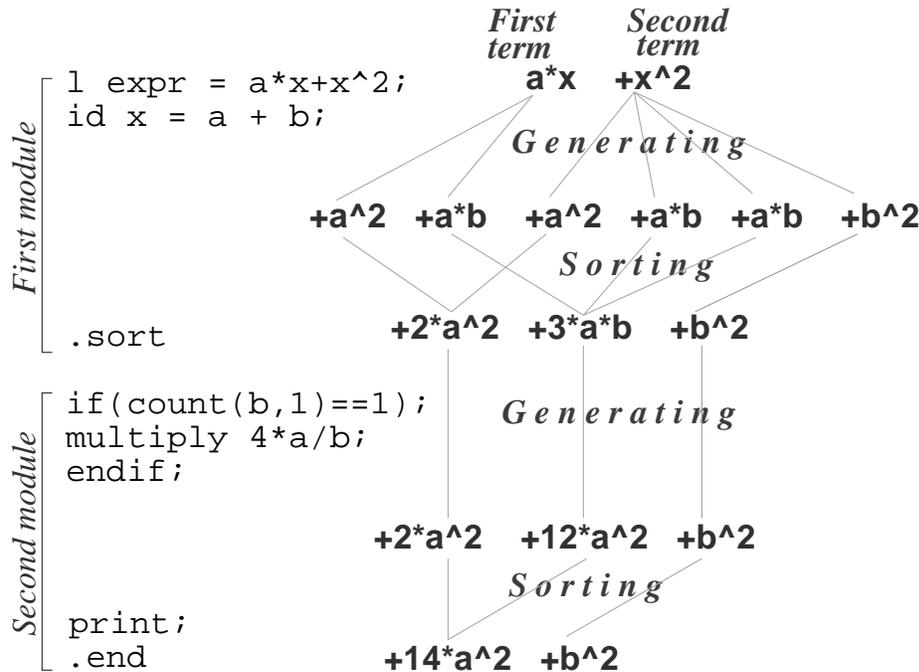}}
\end{center}
\caption{
\label{example}
The fragment of a typical FORM program. In the first module, the
expression $expr = ax+x^2$ is introduced, and then the substitution $x
\to a+b$ is performed. In the second module, only terms in which the
degree of $b$ is exactly 1 are multiplied by $4a/b$ (there is only one such
term in the expression).
}
\end{figure}

FORM is used non-interactively by executing a program that contains
several parts called modules. Modules are terminated with
``dot''-instructions that cause the execution of the module, see the
example on the left of Fig. \ref{example}.

This  example consists of only two modules. There are two
``dot''-instructions: a {\tt .sort} and a {\tt .end}. In both cases the
result is sorted. {\tt .end} additionally terminates the program.

The
execution of each module is divided into three steps:
\begin{itemize}
 \item{\bf Compilation:} The input is translated into an internal
representation.
\item{\bf Generating:} For each term of the input expressions the
statements of the module are executed. This in general generates a lot
of terms for each input term.
\item{\bf Sorting:} All the output terms that have been generated are
sorted and equivalent terms are summed up.
\end{itemize}

FORM only allows local operations on single terms, like replacing parts
of a term or multiplying something to it.
Together with a sophisticated pattern matcher, this at first strong
limitation allows the formulation of general and efficient algorithms.
The limitation to local operations makes it possible to handle expressions
as ``streams'' of terms, that can be read sequentially from a file and
processed independently.

The restriction to local
operations allows to deal with expressions that
are larger than the available main memory and thus in addition allows parallelism.

\section{The Parallelization of FORM}

The limitation of performing only local operations makes FORM very well
suited for parallelization. The concept of parallelization is
straightforward and indicated in Fig. \ref{parexample}:
\begin{figure}[ht]
\begin{center}
\vbox{\epsfxsize=0.82\linewidth \epsfbox{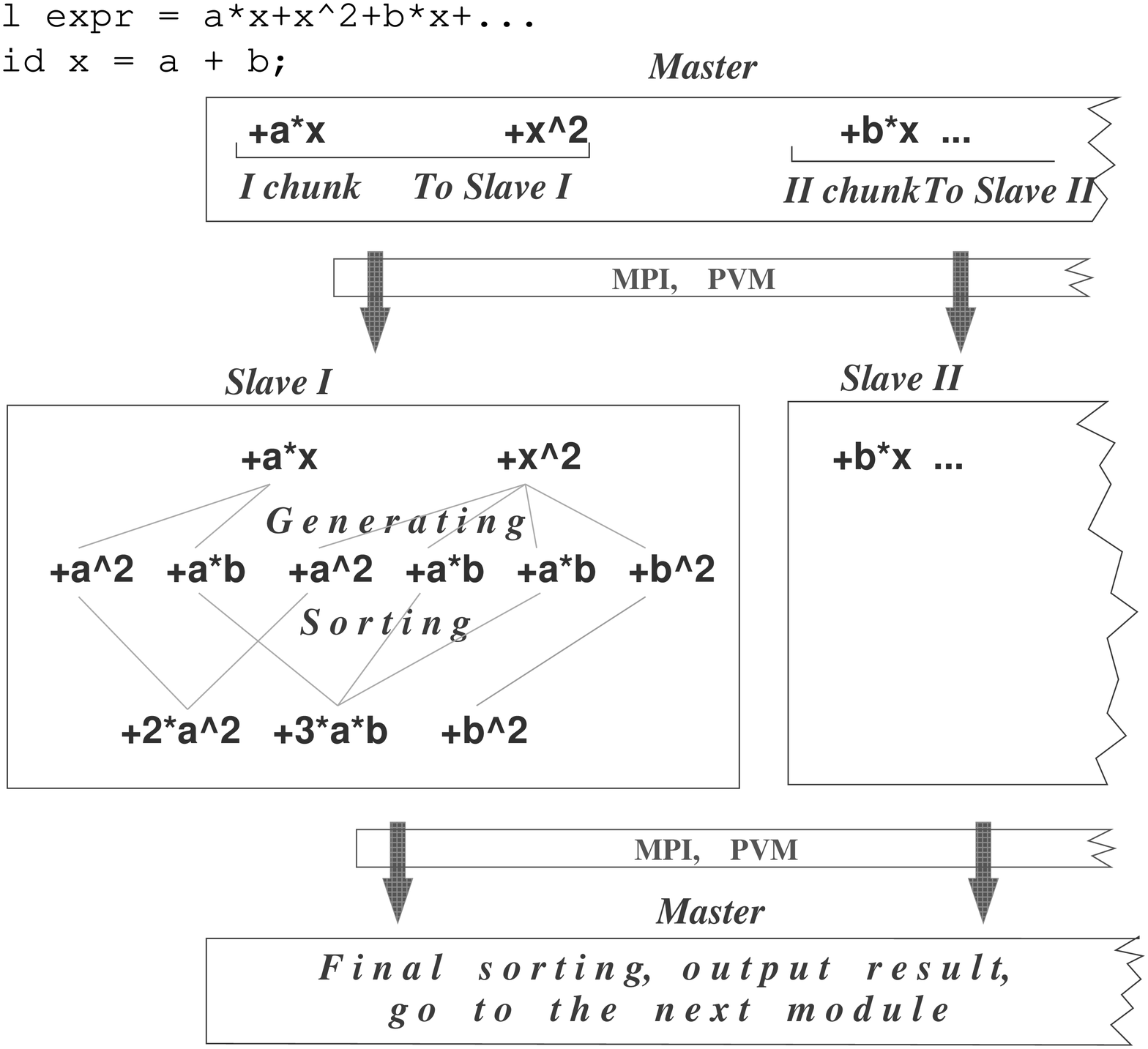}}
\end{center}
\caption{
\label{parexample}
General conception of ParFORM.
}
\end{figure}
Distribute the input terms among available processors, let each of them
perform local operations on its input terms, generate and sort the
arising output terms.
At the end of a module the sorted streams of terms from all processors have
to be  merged to one final output stream again.

This concept indicates to use a master-slave structure for
parallelization. The master would store expressions, distribute
and recollect all terms of each expression.

The master communicates with slaves by means of some message passing
library. Message passing permits to parallelize FORM on computer architectures with
shared or distributed memories, but on the other hand, this leads to some
overhead due to huge data transfers.
Formerly  two libraries were used, MPI and PVM, but
we decided to skip PVM support. The reason was because
many vendors announced to discontinue further development of PVM.
On the other hand, with the announced development of MPI almost all useful PVM
features should appear in MPI.

The master simply distributes and collects data.  With a lower number
of processors, the master becomes almost idle. For that case one can
try to force the master to participate in real calculations, too. On
the other hand, with increasing numbers of slaves, the master spends more
and more time to control slaves, which may lead to early speedup
saturation. Our estimations show that for more than four processors
our Master-Slave model is adequate.

A working parallel FORM prototype ParFORM\cite{Fliegner} was completed in
2000, this was a preliminary version with the syntax of FORM 3, but
without complete FORM 3 features \cite{FORM3}. During the last years, the
real FORM 3 version 3.1 was parallelized. At present, a number of real
physical applications exist which would not have been possible without
ParFORM \cite{PhysAapps}.

As a typical example for physics applications we consider a packet called
``BAICER'' written by P.~Baikov. This is a FORM-packet developed for
reduction of 4-loop propagator massless integrals to some small set of
so-called ``master integrals''. The algorithm is based on recently
developed techniques of solving recurrence relations  using an integral
representation for their solutions \cite{Baikov}. It requires the
calculation of large D expansion coefficients (here ``D'' is the dimension
of the integration space). As a result, the mathematical complexity of the
original problem can be transformed to the necessity to make simple
manipulations but with very large polynomial expressions (billions of
terms and more).

\begin{figure}[ht]
\begin{center}
\vbox{\epsfxsize=0.9\linewidth \epsfbox{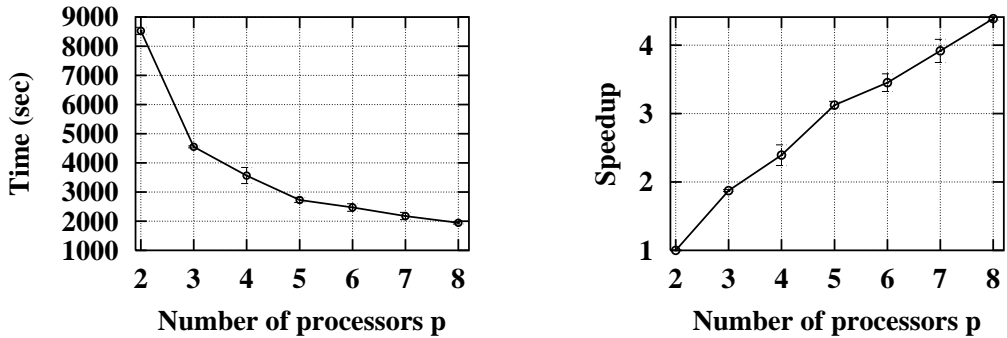}}
\end{center}
\caption{
\label{qcmsmp}
Computing time and speedup for the test program BAICER on
Compaq-AlphaServer with 8x Alpha (EV67) processors 700 MHz.
}
\end{figure}

Both working prototypes of ParFORM and BAICER were developed using the
Karlsruhe Com\-paq-AlphaServer GS60e, 8 Alpha (EV67) processors (700 MHz),
manufactured in 1999. Results for a ``typical'' test of ParFORM using the
BAICER packet are shown in Fig. \ref{qcmsmp}.

The ParFORM structure assumes that almost all real calculations are
performed by slaves while the master only distributes and collects
data. This is the reason why we calculate speedups normalized to the
time spent by programs running on two processors.

Our test program demonstrates an almost linear speedup up to 8
processors available.

From a practical point of view this means that the wall-clock time for
some real tasks can be reduced from months to weeks, which sometimes
is a really essential feature.

Since February 2004 we had a SGI Altix
3000 server available,
and some technical
details will be given:\\
SGI Altix 3700 Server 32x 1.3 GHz/3MB-SC Itanium2 CPUs\\
64 GB DDR/116 MHz mem\\
2.4 TB SCSI disks\\
Red Hat Linux Advanced Server release 2.1AS (Derry).

\begin{figure}[ht]
\begin{center}
\vbox{\epsfxsize=0.9\linewidth \epsfbox{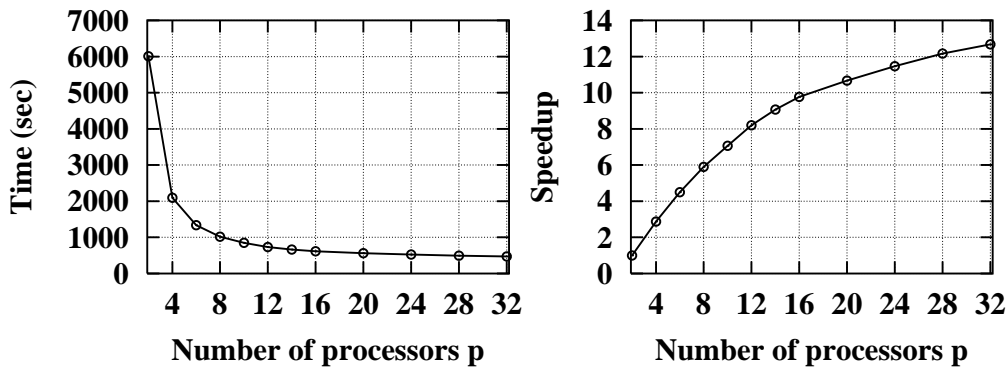}}
\end{center}
\caption{
\label{sgi}
Computing time and speedup for the test program BAICER on the
SGI Altix 3700 server with 32x Itanium2 processors (1.3 GHz).
}
\end{figure}

The results for our test program BAICER are shown in Fig. \ref{sgi}.
The speedup is almost linear up to 12 processors.
Of course hereafter the speedup is not linear, but it is still
considerable.

An achieved speedup of 12 means that a FORM job that would need one year of
computing time can be run as ParFORM job in less than one month. This leads
physics to a qualitatively new level, because it would practically be
impossible to run jobs for years whereas months are feasible nowadays.

Fig. \ref{sgi} shows that with 16 processors we have a speedup of 10. This
means that we can run on our 32-processor computer two jobs
simultaneously, having the speedup of 10 for each of them.

\section{Conclusion}
We have shown that the internal structure of FORM permits a
``natural'' parallelization of parallelizable problems.
It is worth mentioning that these results have been achieved with FORM
programs that were written
for the sequential version and have not been modified. Generally the
FORM user does not have to know anything about the mechanism behind
the
parallel version to run existing programs in parallel. Still,
some knowledge can help to tune them and achieve a higher speedup.

Of course the speedup that can be achieved strongly depends on the
problem under consideration.

In some cases, i.e. ideal FORM input code and adequate problem size, the
achieved speedup is almost linear in the number of slave processors. For
realistic complex applications, the speedup is still considerable even
with a larger number of processors.

Colleagues who are interested to use ParFORM should contact
M.~Tentyukov
by e-mail\\  tentukov@particle.uni-karlsruhe.de

We thank Prof. Dr. J. K\"uhn for interesting discussions and
encouragement and the DFG-SFB-TR09 project for support.

\end{document}